\documentclass[aps,nofootinbib,amsmath,prd,onecolumn,notitlepage,showpacs,superscriptaddress,11pt]{revtex4-1}

\usepackage[american]{babel}
\usepackage{amsfonts}
\usepackage{amsmath}
\usepackage{amssymb}
\usepackage{slashed}
\usepackage{xcolor}
\usepackage{bm}
\usepackage{float}
\usepackage[utf8]{inputenc}
\usepackage{nicefrac}
\usepackage{hyperref}
\usepackage{url}
\usepackage{tikz-cd}
\usepackage[normalem]{ulem}

\definecolor{darkblue}{rgb}{0.1,0.1,0.7}
\hypersetup{colorlinks,
           linkcolor={darkblue},
           citecolor={darkblue},
           urlcolor={darkblue}
}

\newcommand{\lvec}[2]{\raise #1\hbox{$^\leftarrow$} \hspace{-9pt} #2}
\newcommand{\rvec}[2]{\raise #1\hbox{$^\rightarrow$} \hspace{-9pt} #2}
\newcommand{\lrvec}[2]{\raise #1\hbox{$^\leftrightarrow$} \hspace{-9pt} #2}

\newcommand{\dd}{{\rm d}}

\usepackage[normalem]{ulem}

\allowdisplaybreaks[4]
\begin{document}

\title{Euclidean actions and static black hole entropy in teleparallel theories}

\author{Iberê Kuntz}
\email{kuntz@fisica.ufpr.br}
\affiliation{
Departamento de F\'isica, Universidade Federal do Paran\'a, PO Box 19044, Curitiba -- PR, 81531-980, Brazil}

\author{Gregorio Paci}
\email{gregorio.paci@phd.unipi.it (Corresponding Author)}
\affiliation{
Universit\`a di Pisa and INFN -- Sezione di Pisa, Largo Bruno Pontecorvo 3, 56127 Pisa, Italy}

\author{Omar Zanusso}
\email{omar.zanusso@unipi.it}
\affiliation{
Universit\`a di Pisa and INFN -- Sezione di Pisa, Largo Bruno Pontecorvo 3, 56127 Pisa, Italy}

\begin{abstract}
%
\noindent It is well-known that the results by Bekenstein, Gibbons and Hawking on the thermodynamics of black holes can be reproduced quite simply in the Euclidean path integral approach to Quantum Gravity. The corresponding partition function is obtained semiclassically, ultimately requiring only the on-shell Einstein--Hilbert action with opportune asymptotic subtractions. 
We elaborate on the fact that the same expressions for the thermodynamical quantities can be obtained within teleparallel equivalent theories, based on either torsion or nonmetricity, by employing quasilocal relations. Notably, the bulk integrals of these theories do not vanish on-shell but rather result in boundary terms themselves. Asymptotic subtractions of the latter are able to cancel out the divergences, ultimately leading to Bekenstein--Gibbons--Hawking's results.
As a non-trivial cross-check, we compute the bulk integrals directly without reference to the boundary terms. While the result agrees with the previous method for the torsion-based teleparallel theory, it differs for the nonmetricity theory. Specifically, upon regularizing the bulk integral using a fiducial reference frame, we find that the semiclassical partition function vanishes.
To address this problem, we propose a simple prescription for Schwarzschild black holes, which involves keeping the nonmetric connection arbitrary and imposing thermal equilibrium.
Generalizations of the results to more general modified gravity theories with antisymmetric degrees of freedom are also discussed.
\end{abstract}

\pacs{}
\maketitle

\section{Introduction}\label{sect:intro}

Hawking’s discovery that quantum effects cause black holes to radiate particles has put black hole thermodynamics on a solid foundation, thus showing that it is more than a mere formal analogy. This result is particularly striking from a classical perspective, according to which black holes would be expected to have infinite entropy, due to the no-hair theorem, and zero temperature, since nothing can escape them. Consequently, the challenge of rigorously defining thermodynamic quantities, such as energy, in black hole physics becomes even more pressing.

In the standard formulation of General Relativity (GR) we cannot have a definition of energy of the gravitational field which is, at the same time, local and covariant.
The complete action of General Relativity (GR) over a finite region ${\cal M}$ of spacetime with boundary $\partial{\cal M}$ and Dirichlet boundary conditions reads:
\begin{equation}\label{eq:gr-action-lorentzian}
 \begin{split}
  S_{GR} [g] = \frac{1}{16\pi} \int_{\cal M} \dd^4 x \sqrt{-g} \mathring{R}
  + \frac{1}{8\pi} \int_{\partial {\cal M}} \dd^3 y  \, \varepsilon \Bigl(
  \sqrt{|h|} K -\sqrt{|h{}^{(0)}|} K{}^{(0)} \Bigr)
  \,,
 \end{split}
\end{equation}
where we introduced the Ricci scalar curvature $\mathring{R}$, the induced metric on the boundary $h_{ab}$, and the extrinsic curvature $K_{ab}$, which corresponds to the Gibbons--Hawking--York boundary (GHY) term \cite{York:1972sj,Gibbons:1976ue}.
The trace of the extrinsic curvature $K$ on the boundary, which is multiplied by the coefficient $\varepsilon=\pm1$ for timelike/spacelike regions of $\partial{\cal M}$, is required for a proper cancellation of boundary normal derivatives when varying the action. The action also includes a nondynamical subtraction based on a boundary metric $h^{(0)}_{ab}$, which is isometric to $h_{ab}$, but assumed to be embedded in a reference spacetime (e.g., flat spacetime). The subtraction term is not required for obtaining the right dynamics, but, rather, it is needed for the finiteness of several physical quantities as we shall see.

Using the Arnowitt--Deser--Misner (ADM) $3+1$ formalism for some ${\cal M}$ with the topology of a foliated cylinder, it is possible to compute the Hamiltonian associated with $S_{GR} [g]$ and thus provide a notion of energy \cite{Arnowitt:1962hi}. Spacetime is foliated in terms of spacelike hypersurfaces $\Sigma_t$, with $t\in[t_1,t_2]$, which intersect the timelike part of the boundary $\partial{\cal M}$ in codimension-two surfaces $B_t= \Sigma_t \cap {\partial {\cal M}}$ for $t\in(t_1,t_2)$. The surfaces $B_t$ can be seen as the celestial spheres in asymptotically flat spacetimes with metric $\sigma_{AB}$.
The Hamiltonian includes a bulk and a boundary term, just like $S_{GR}[g]$, but, on-shell, only the boundary term survives and gives a notion of ADM mass/energy
\begin{equation}
 \begin{split}
  M_{\rm ADM} = -\frac{1}{8\pi}\int_{B_t} ( \kappa -\kappa_0) \sqrt{\sigma} \dd^2 \theta
  \,,
 \end{split}
\end{equation}
where ${\kappa}_{ab}$ is the extrinsic curvature of $B_t$ as embedded in the $\Sigma_t$ hypersurfaces. In essence, the above energy can be seen as a foliation of the (timelike part of the) boundary terms of $S_{GR}[g]$. We stress that this energy is given entirely by a boundary integral,
which is occasionally referred to as a quasilocal definition of mass/energy \cite{Brown:1992br}. Other commonly used definitions of mass include the Komar and Bondi-Sachs masses, both of which are expressed as boundary integrals, similar to $  M_{\rm ADM}$. 
However, for the static and spherically symmetric black holes that we will consider, it is known that these definitions coincide with the ADM mass. Thus, we will not explicitly discuss them here.

An alternative route to derive $M_{\rm ADM}$ involves the use of pseudotensors,
with which it is possible to provide the notion of gravitational energy-momentum ``pseudotensor'', but this comes at the expense of the loss of covariance.
Following Weinberg \cite{Weinberg:1972kfs}, we assume the existence of an asymptotically flat frame and introduce the Lorentz-tensor $h_{\mu\nu}$ implicitly as $g_{\mu\nu} = \eta_{\mu\nu} + h_{\mu\nu}$, whose indices are raised and lowered with the flat metric $\eta_{\mu\nu}$ rather than with $g_{\mu\nu}$.\footnote{%
The alternative approach by Landau and Lifshitz gives the same results in the end \cite{Landau:1975pou}.
} Vacuum Einstein's equations can then be rewritten as
\begin{equation}
 \begin{split}
  -\partial^2 h_{\mu\nu} + 2\partial_\alpha \partial_{(\mu}  h^{\alpha}{}_{\nu)}-\eta_{\mu\nu} \partial_\alpha\partial_\beta h^{\alpha\beta} + \eta_{\mu\nu} \partial^2 h^{\alpha}{}_\alpha
  -\partial_\mu \partial_\nu h^{\alpha}{}_\alpha
  =
  16\pi t^{\rm W}_{\mu\nu} \,,
 \end{split}
\end{equation}
where $t^{\rm W}_{\mu\nu}\sim O(h^2)$ is Weinberg's gravitational energy-momentum pseudotensor. It can be used to give local but noncovariant notions of energy, momentum and angular momentum. In particular, the total energy of a region comes from integrating the $t^{\rm W}_{00}$
component. Using Stokes' theorem and the equations of motion, the volumetric/bulk integral can be rewritten as a boundary/surface one:
\begin{equation}\label{eq:weinberg-energy}
 \begin{split}
  M_{\rm W} = \frac{1}{16\pi}\int_{S_r \to \infty} (D^i h_{ij} -D_j h^i{}_i) n^j \sqrt{\sigma} \dd^2 \theta
 \end{split}
 \ ,
\end{equation}
where $n^i$, with $i=1,2,3$, is the normal to $S_r$, which is the large-$r$ hypersurface at constant $t$ in the rest frame with metric $\sigma_{ab}$, and coordinates $\theta^a$ with $a=1,2$. Given an asymptotically flat spacetime and identifying the boundaries $B_t$ and $S_r$ of the two approaches, it is not difficult to show that $M_{\rm ADM}= M_{\rm W}$.

Up to now, the main takeaway message of this introduction is that the physical observable, i.e., the energy,
is given not as a local density integrated over a volume, but rather as a nonlocal integral over the boundary of a region.
Weinberg's manipulation is formal, in the sense that the volumetric integral is never truly evaluated, and, in fact, the components of $h_{\mu\nu}$ can be highly singular inside the region.
Importantly, the manipulations work by virtue of two ingredients: the equations of motion and Stokes' theorem.

As we shall review briefly in the next section, the same considerations hold when the Euclidean action of GR is evaluated on-shell. The Euclidean action of GR is proportional to the ``free energy'' of the gravitational system, e.g., a black-hole, and its evaluation subsequently yields its thermodynamical properties through standard manipulations \cite{Bardeen:1973gs}.
In this sense, both the energy and the entropy have a quasilocal origin
since they are both determined from the Euclidean action, which on-shell
involves only boundary terms when matter is absent \cite{York:1986it}.

Let us now point out that the metric-based action \eqref{eq:gr-action-lorentzian} is not the unique description of gravity, but there are equivalent descriptions based on either torsion or nonmetricity degrees of freedom, known as teleparallel equivalent theories \cite{Aldrovandi:2013wha}.
The equivalence of these theories to GR holds at level of the equations of motion, but otherwise their actions contain boundary terms not present in the Einstein-Hilbert one. Such boundary terms, albeit having no effect at the classical level, cannot be neglected in the path integral.  In the first part of this paper, we will focus specifically on these terms.
In particular, we will show that by applying the same steps outlined above--namely, formally using Stokes' theorem and evaluating the observables on-shell--the mass/energy and entropy are always given in terms of quasilocal integrals, 
even in teleparallel equivalent theories. This is somewhat surprising
because teleparallel theories are formulated in terms of rather different fields (torsion or nonmetricity) \cite{BeltranJimenez:2019esp}, even though they propagate the same degrees of freedom, and they do not require boundary terms for dynamical purposes \cite{Erdmenger:2022nhz,Erdmenger:2023hne}. We show that nondynamical subtraction terms are needed, but they are as natural as the subtraction of the GHY term.

Interestingly, unlike in GR, where the bulk contribution vanishes, the Euclidean actions of teleparallel theories can be evaluated as volumetric integrals. Although bulk integrals must ultimately yield the same results as boundary integrals, they require quite \emph{different} regularization procedures to be well-defined. As a result, showing that these two approaches lead to identical outcomes is not immediately obvious, but instead it constitutes an informative and non-trivial cross-check of the thermodynamic properties obtained. This is our primary motivation for examining bulk integrals as well.
To achieve this result, we shall adopt a prescription proposed in Ref.~\cite{Krssak:2015rqa} which--when applied to the volumetric integrals--is analogous to pseudotensors in GR, as it requires additional reference structures. We shall discuss this point in more detail later. Surprisingly, this approach shall yield different results depending on the type of teleparallel model. In particular, in the case of the nonmetricity-based theory, we will find a vanishing partition function. A similar outcome, using a different methods, were reported in Ref.~\cite{Krssak:2024kva,BeltranJimenez:2018vdo}, where a doubling of the GR result was obtained. A rather formal and general solution to this conundrum based on the Wald entropy was proposed in Ref.~\cite{Heisenberg:2022nvs}. Here, we introduce a simpler and more practical prescription which also yields the desired result for Schwarzschild.

The paper is structured in such a way that we address first the quasilocal formulations in Sect.~\ref{sect:quasilocal}, and then the volumetric ones in Sect.~\ref{sect:bulk}. A brief review of the metric case is given for setting up conventions and normalizations. In Sect.~\ref{sect:corrections} we discuss an extension of some results that may be useful for modified theories of gravity based on the antisymmetric teleparallel framework.

\section{Quasilocal framework}\label{sect:quasilocal}

In this section we confirm that (Euclidean) teleparallel theories equivalent to GR
give exactly the same quasilocal formulas for energy and entropy of a static spherically-symmetric Schwarzschild black hole
under the \emph{same} set of basic assumptions. We begin by reviewing the standard case of GR for reference.

\subsection{Brief review of the Euclidean action in GR}\label{sect:quasilocal-GR}

The complete Euclidean counterpart to the action \eqref{eq:gr-action-lorentzian} of GR over a region ${\cal M}$ of spacetime is
\begin{equation}\label{eq:gr-action-euclidean}
 \begin{split}
  S_E [g] = -\frac{1}{16\pi} \int_{\cal M} \dd^4 x \sqrt{g} \mathring{R}[g]
  -\frac{1}{8\pi} \int_{\partial {\cal M}} \dd^3 y \sqrt{h} K
  +\frac{1}{8\pi} \int_{\partial {\cal M}} \dd^3 y \sqrt{h^{(0)}} K^{(0)}\,.
 \end{split}
\end{equation}
Given the uniformly positive signature of the Euclidean case, the boundary has also fixed signature, i.e., the parameter $\varepsilon=1$ as introduced in \eqref{eq:gr-action-lorentzian} is fixed. The Euclidean action is normalized in such a way that $i S_{GR} = - S_{E}$ after a Wick rotation,
so that the Euclidean partition function can be evaluated semiclassically
\begin{equation}\label{eq:gr-pi-euclidean}
 \begin{split}
  Z = \int {D}g \, {\rm e}^{- S_{E}} \approx {\rm e}^{- S_{E}[g]}
 \end{split}
\end{equation}
from the path-integral using a dominant field configuration, which is denoted again $g_{\mu\nu}$ for notational simplicity. Then $S_{E}[g]$ becomes essentially $\beta$ times the free energy, where $\beta$ is the inverse temperature.
As dominant contribution to the path-integral we take the Euclidean version of (the exterior of) a Schwarzschild black holes. The Schwarzschild metric solves Einstein's equations in vacuum, thus $\mathring{R}=0$. We refer to the appendix \ref{appendix:euclidean-schwarzschild} for more details on the geometry of the Euclidean black hole.

For the computation of the semiclassical path-integral,
we must integrate the action in the range $r_s \leq r <r_0$ of the Euclidean Schwarzschild patch for $r_s=2m$, which defines the region ${\cal M}$, and then send $r_0\to \infty$.
For $\beta = \beta_H = 8\pi m$ the Euclidean patch is smooth everywhere, including at $r=2m$ using for example Eddington-Finkelstein or Kruskal-Szekeres's coordinates. By construction, we have that the boundary $\partial {\cal M}$
coincides with the hypersurface $r=r_0$, and there is no inner boundary. The boundary has the topology $\partial{\cal M}\simeq S^1 \times S^2$ (periodic time and celestial sphere) and unit normal vector $n_\mu = f^{-1/2} \partial_\mu r$.
We rewrite the metric similarly to the standard $3+1$ decomposition using the hypersurfaces at constant-$r$ as foliation
\begin{equation}
 \begin{split}
  \dd s^2|_E = N^2 \dd r^2 + h_{ab} \dd y^a \dd y^b \,,
 \end{split}
\end{equation}
where $N= f^{-1/2}$ plays the role an Euclidean lapse-function and $h_{ab} \dd y^a \dd y^b=f(r) \dd \tau^2 + r^2 \dd \Omega^2$ is the metric on a constant-$r$ hypersurface.
The extrinsic curvature is related to the Lie derivative with respect to $r^\mu$ of the metric $h_{ab}$ on the hypersurfaces. Knowing that
$\partial_r = r^\mu \partial_\mu$ and $r^\mu = N n^\mu$, we have the standard relation
\begin{equation}
 \begin{split}
 {\cal L}_{r^\mu} h_{ab} = 2 N K_{ab} = 2 f(r)^{-1/2} K_{ab} \,,
 \end{split}
\end{equation}
but in these coordinates ${\cal L}_{r^\mu} h_{ab} = \partial_r h_{ab}$,
which implies
\begin{equation}
 \begin{split}
 K= h^{ab} K_{ab} = \frac{1}{2}f(r)^{1/2} h^{ab} \partial_r h_{ab}
 = \frac{1}{2}f(r)^{1/2}(f(r)^{-1} f'(r) + 4/r)
 \end{split}
\end{equation}
after substitution of the components.
We also have the volume element on the boundary
$
 \sqrt{h} \dd ^3 y = f(r)^{1/2} r^2 \dd \tau \dd^2 \Omega
$.
Combining everything together we have that the GHY term gives
\begin{equation}
 \begin{split}
 -\frac{1}{8\pi} \int_{\partial {\cal M}} \dd^3 y \sqrt{h} K
 &= \beta \Bigl(-r_0 + \frac{3}{4} r_s\Bigr) + O(r_0^{-1})\,,
 \end{split}
\end{equation}
which diverges for $r_0\to \infty$, and the divergence must be cured by the subtraction.

The computation of the subtraction term is more delicate.
The metric $h{}^{(0)}_{ab}$ that has to have the same \emph{intrinsic} line element
\begin{equation}
 \begin{split}
 h^{(0)}_{ab}\dd y^a \dd y^b = h_{ab}\dd y^a \dd y^b
 \end{split}
\end{equation}
at $r=r_0$, but it must come from the embedding on a flat $4d$ metric.
For consistency, the flat embedding must also be at finite temperature, obviously rescaled by the correct radius of the $S^1$ with coordinate $\tau$ at $r=r_0$.
All these requirements are accomplished by choosing
\begin{equation}
 \begin{split}
 h^{(0)}_{ab}\dd y^a \dd y^b = f_0 \dd \tau^2 + r^2 \dd \Omega^2 \,,
 \end{split}
\end{equation}
where $f_0= f(r_0)$, and the embedding is 
$
  \dd s^2|_{E,{\rm flat}} = \dd r^2 + h^{(0)}_{ab} \dd y^a \dd y^b 
$.
The normal vector is now $n^{(0)}_\mu =\partial_\mu r$, so $N^{(0)}=1$, but, other than this, the same steps of the previous computation apply.
Using the new embedding the main differences are that $f_0$ will not vary with $r$ and we do not have an $r$-dependent normalization on $K^{(0)}$ caused by $N^{(0)}=1$, so 
\begin{equation}
 \begin{split}
 K^{(0)}= \frac{2}{r} \,.
 \end{split}
\end{equation}
We also have the volume element
$
 \sqrt{h^{(0)}} \dd ^3 y = f_0^{1/2} \dd \tau r^2 \dd^2 \Omega 
$.
Again, combining everything together we have that, for $f_0=f(r_0)$, subtraction term is
\begin{equation}
 \begin{split}
 -\frac{1}{8\pi} \int_{\partial {\cal M}} \dd^3 y \sqrt{h^{(0)}} K^{(0)}
 &= \beta \Bigl(-r_0 + \frac{1}{2} r_s\Bigr) + O(r_0^{-1})
 \,,
 \end{split}
\end{equation}
whose divergence in $r_0\to \infty$ is precisely shaped to cancel the one of the GHY term.
The integral of the scalar curvature is given in the appendix and it is nonzero only if $\beta \neq \beta_H$ through a potential conical singularity.

\subsection{Thermodynamics}\label{sect:thermodynamics}

We begin at equilibrium, i.e., $\beta = \beta_H = 8\pi m = 4\pi r_s$ \cite{York:1986it}.
Combining the two terms computed in the previous part, we find that the action on-shell is finite in the limit $r_0\to \infty$
\begin{equation}
 \begin{split}
  S_E [g] = \beta \frac{r_s}{4} = \frac{\beta^2}{16\pi}
 \end{split}
\end{equation}
without bulk contributions (recall that $\mathring{R}=0$ for the Schwarzschild metric at $\beta= \beta_H$).
The dominant part of the semiclassical partition function is
\begin{equation}
 \begin{split}
  Z(\beta)= \exp(-S_E [g]) = \exp\Bigl(-\frac{\beta^2}{16\pi}\Bigr)\,,
 \end{split}
\end{equation}
from which we can compute the thermodynamical energy as
\begin{equation}
 \begin{split}
  E = -\frac{\partial}{\partial \beta} \log Z(\beta)= \frac{\beta}{8\pi} = m
  \,,
 \end{split}
\end{equation}
coinciding with the mass of the black hole.
The entropy is derived as
\begin{equation}
 \begin{split}
  S = \Bigl(1-\beta\frac{\partial}{\partial \beta}\Bigr) \log Z(\beta)= \frac{\beta^2}{16\pi} = \pi r_s^2 = \frac{A}{4}
  \,,
 \end{split}
\end{equation}
where $A=4\pi r_s^2 = \pi m^2$ is the area of the event horizon. This is a manifestation of Bekenstein-Hawking's celabrated result \cite{Bekenstein:1972tm,Hawking:1975vcx}. It is important to realize that
up to now all quantities depend only on one energy scale, that is, the scale set by $\beta$. This happens because we required a smooth Euclidean patch \cite{Gibbons:1976ue}.

Outside equilibrium, we have that in general $\beta \neq \beta_H$, which implies the presence of a conical singularity at $r=r_s=2m$. Now $\beta$ and $m$ are two independent quantities.
It is convenient to define the canonical energy in presence of the singularity
as
\begin{equation}
 \begin{split}
  \beta F(m,\beta) = S_E[g] \,.
 \end{split}
\end{equation}
The boundary terms evaluated in the previous section do not change as they have been evaluated at $r_0\to \infty$, which is far away from the conical singularity, hence:
\begin{equation}
 \begin{split}
 -\frac{1}{8\pi} \int_{\partial {\cal M}} \dd^3 y \sqrt{h} K+\frac{1}{8\pi} \int_{\partial {\cal M}} \dd^3 y \sqrt{h^{(0)}} K^{(0)}
 &= \frac{\beta r_s}{4}=\frac{\beta m}{2}
 \,.
 \end{split}
\end{equation}
However, there is no longer a relation between $\beta$ and $m$.
On the other hand, we now have
the contribution from the bulk integral on ${\cal M}$
over an arbitrarily small region containing the singularity:
\begin{equation}
 \begin{split}
  \int_{\rm sing} \dd^4 x \sqrt{g} R 
  =
  4\pi \Bigl( 1- \frac{\beta}{\beta_H}\Bigr) A
  \,.
 \end{split}
\end{equation}
Combining now bulk and boundary we have the canonical energy
\begin{equation}
 \begin{split}
  \beta F(m,\beta) = \frac{\beta m}{2} -\frac{1}{16\pi}\Bigl( 1- \frac{\beta}{8\pi m}\Bigr) A
  = \frac{\beta m}{2}- 4\pi m^2 \Bigl( 1- \frac{\beta}{8\pi m}\Bigr)
  \,.
 \end{split}
\end{equation}
The condition for equilibrium is that $\frac{\partial F}{\partial m}=0$,
from which we can determine the equilibrium temperature. Using the above formula
\begin{equation}
 \begin{split}
  \beta \frac{\partial F}{\partial m} = \beta -8\pi m = \beta -\beta_H
 \end{split}
\end{equation}
so $\beta =\beta_H$, that is Hawking's temperature, is actually the equilibrium temperature. The equilibrium temperature is thus also the temperature for which there is no conical singularity. Using $\beta=\beta_H$ and $Z=\exp(-\beta F)|_{\beta=\beta_H}$
at equilibrium we recover the same thermodynamics discussed at the beginning of this section.

\subsection{Euclidean action in TEGR}\label{sect:quasilocal-TEGR}

The requirement that the connection of GR is Levi-Civita's is kinematical, rather than dynamical. An alternative construction can be achieved by requiring that the connection, say $\nabla$, is flat and metric-compatible, but has torsion.
The result of this alternative kinematical constraint is known as (antisymmetric) teleparallelism. For any two vectors $v$ and $w$, torsion is defined 
\begin{equation}
 \begin{split}
  T(v,w) = \nabla_v w-\nabla_w v -[v,w]\,,
 \end{split}
\end{equation}
and its components can be expressed in any basis, e.g., $T(v,w) =T^{\alpha}{}_{\beta\gamma} v^\beta w^\gamma \partial_\alpha$.
The connection $\nabla$ can always be recasted as the sum of Levi-Civita's $\mathring{\nabla}$ and a contortion part that depends on the torsion components, so that their curvatures can be related as well. With a bit of work, it is straightforward to show that requirement of teleparallelism implies
\begin{equation}\label{eq:tegr-teleparallelism}
 \begin{split}
  0=g^{\beta\nu} R[\nabla]^\alpha{}_{\beta \alpha \nu}
  =  \mathring{R} + \mathring{\mathbb{T}} + 2 \mathring{\nabla}_\mu T^\mu\,,
 \end{split}
\end{equation}
where $T_\mu=T^\nu_{\ \mu \nu}$ is known as torsion vector, and we defined the torsion-scalar
\begin{equation}
 \begin{split}
  \mathring{\mathbb{T}} = -\frac{1}{4} T_{\alpha\mu\nu}T^{\alpha\mu\nu}
  -\frac{1}{2}T_{\alpha\mu\nu}T^{\mu\alpha\nu}
  + T_\mu T^\mu\,.
 \end{split}
\end{equation}
In practice, the scalar curvature $\mathring{R}$ differs from the scalar $\mathring{\mathbb{T}}$ only by a total derivative.
Having set all the necessary ingredients, the complete teleparallel equivalent to GR (TEGR) action is
\begin{equation}\label{eq:tegr-action-euclidean}
 \begin{split}
  S_{T} [g,\nabla] = \frac{1}{16\pi} \int_{\cal M} \dd^4 x \sqrt{g} \, \mathring{\mathbb{T}}
  + S_{\rm sub}
  \,,
 \end{split}
\end{equation}
where $\nabla$ is compatible with $g_{\mu\nu}$, but flat and torsionful.
We have also included a nondynamical subtraction term $S_{\rm sub}$ in analogy to the case of GR, which we determine in a moment with an educated guess as in \cite{Oshita:2017nhn}.
Using Eq.~\eqref{eq:tegr-teleparallelism} and up to the term $S_{\rm sub}$, we observe that $S_{T}$ is quasi-equivalent to the Einstein--Hilbert action $S_E[g]$, as they differ only by the boundary term $2 \mathring{\nabla}_\mu T^\mu$, which does not affect the equations of motion. Importantly, this boundary term need not to be introduced \textit{a posteriori} as it happens for the Gibbons--Hawking--York term. Moreover, Eq.~\eqref{eq:tegr-action-euclidean} is first order in derivatives, thus precluding the need of the latter.
However, it should be emphasized that the aforementioned equivalence is not guaranteed to hold at the quantum level because boundary terms do play a role in general.

Recall that the constraints on the connection are kinematical, rather than dynamical. The teleparallel connection can thus be solved in terms of a local $GL(4)$ matrix satisfying the additional integrability conditions of compatibility. The simplest solution comes from introducing a local basis of vierbeins, $g_{\mu\nu}=e^a{}_\mu e^b{}_\nu \eta_{ab}$ (notice that $\eta_{ab}=\delta_{ab}$ in the Euclidean case) and requiring that their spin-connection is zero.\footnote{%
Other choices are possible, as we shall see later.
} In this case then the constraints on the connection are solved in terms of a Weitzenb\"ock connection, i.e., $\nabla$ has components
\begin{equation}
 \begin{split}
  \Gamma^\alpha{}_{\beta \mu} = E^\alpha{}_a \partial_\mu e^a{}_\beta
  \,,
 \end{split}
\end{equation}
where $e^a{}_\mu$ are the vierbeins and $E^\alpha{}_a$ are the inverse vierbeins.
The torsion is thus
\begin{equation}
 \begin{split}
  T^\alpha{}_{ \mu \beta} = E^\alpha{}_a (\partial_\mu e^a{}_\beta - \partial_\beta e^a{}_\mu)
 \end{split}
\end{equation}
in holonomic components.
Inserting this expression in $S_T$ given in \eqref{eq:tegr-action-euclidean} makes the action a functional of the vierbeins only. It also reveals that
boundary terms are not needed because $\mathring{\mathbb{T}}$ depends at most on one derivative of the now dynamical vierbeins, even though the action is nonlinear because of $E^\alpha{}_a$.\footnote{%
For a general argument based on a stronger requirement of teleparallelism see \cite{Erdmenger:2023hne}.
} A double check is that the variation of $S_{T}[e]$ reproduces Einstein's equations, but with the metric expressed in terms of the vierbeins (the antisymmetric part of the equations is automatically zero).

For the computation of the Euclidean action on-shell, recall the requirement of teleparallelism \eqref{eq:tegr-teleparallelism} once more. On-shell $S_T$ gives Einstein's equations in vacuum, thus $\mathring{R}=0$
for vacuum solutions, which, in the teleparallel theory, imples also $\mathring{\mathbb{T}} =- 2 \mathring{\nabla}_\mu T^\mu$.
The torsion scalar can thus be rewritten as a total divergence, and then as a boundary term on-shell using Stokes' theorem formally.
Consequently, we have for the Euclidean action on-shell
\begin{equation}
 \begin{split}
  S_{T} [e]
  = -\frac{1}{8\pi} \int_{\partial{\cal M}} \dd^3 y \sqrt{h} \, n_\mu T^\mu
  + S_{\rm sub}
 \end{split}
\end{equation}
assuming as in the case of GR only a boundary at large $r$. We have all the ingredients to formulate an educated guess for the subtraction term, for which we follow the same logic as in the case of GR. We construct it with quantities coming from an ``isometric'' embedding in asymptotically flat space
\begin{equation}
 \begin{split}
  S_{T} [e]
  = -\frac{1}{8\pi} \int_{\partial{\cal M}} \dd^3 y \sqrt{h} \, n_\mu T^\mu
  + \frac{1}{8\pi} \int_{\partial{\cal M}} 
  \dd^3 y \sqrt{h^{(0)}} \, n^{(0)}_\mu T_{(0)}{}^\mu
  \,,
 \end{split}
\end{equation}
which we clarify in a moment. The choice coincides with the one made in Refs.~\cite{Oshita:2017nhn,BeltranJimenez:2018vdo}.

For the explicit evaluation we use the simplest diagonal frame $e^a{}_\mu = {\rm diag}(f^{1/2},f^{-1/2},r, r\sin \theta)$, which gives the Euclidean Schwarzschild metric in the contraction with itself. The normal vector to constant-$r$ hypersurfaces is $n^\mu = (0,f^{1/2},0,0)$ and the determinant of the hypersurfaces metric is $\sqrt{h}\dd^3 y= r^2 f^{1/2} \dd\tau \dd^2\Omega$. This gives the boundary term as
\begin{equation}
 \begin{split}
  -\frac{1}{8\pi} \int_{\partial{\cal M}} \dd^3 y \sqrt{h} \, n_\mu T^\mu
  = -\frac{1}{8\pi} \int r^2 f(r) \dd\tau \dd^2\Omega T^\nu{}_{r\nu}
  \,.
 \end{split}
\end{equation}
We need only the radial component of the torsion vector $T_\mu=T^\nu{}_{\mu\nu}$, that is,
\begin{equation}
 \begin{split}
  T^\nu{}_{r\nu} &= E^\nu{}_a \partial_r e^a{}_\nu -E^\nu{}_a \partial_\nu e^a{}_r=
 \frac{1}{2}(f^{-1} f' + 4/r)
 \,.
 \end{split}
\end{equation}
Inserting everything in the integral we find exactly the same result as the GHY boundary term in GR at $r=r_0$,
\begin{equation}\label{eq:tegr-boundary}
 \begin{split}
 -\frac{1}{8\pi} \int_{\partial{\cal M}} \dd^3 y \sqrt{h} \, n_\mu T^\mu
 &= \beta \Bigl(-r_0 + \frac{3}{4} r_s\Bigr) + O(r_0^{-1})\,.
 \end{split}
\end{equation}
which retroactively justifies the need for a subtraction to become finite in the limit $r_0\to \infty$.

For the subtraction we follow once more the same strategy as in GR.
The embedding vierbein is chosen such that
$e^{(0)}{}^a{}_\mu \dd x^\mu=e^a{}_\mu \dd x^\mu
$
only the hypersurface $r=r_0$, and it is chosen as a vierbein of a flat metric, i.e., ${\rm diag}(f_0,1,r,r\sin \theta)$, in which $\tau$ is an angular variable.
The obvious choice is $e^{(0)}{}^a{}_\mu = {\rm diag}(f_0^{1/2},1,r, r\sin \theta)$.
In this case $\sqrt{h^{(0)}}\dd^3 y= r^2 f_0^{1/2} \dd\tau \dd^2\Omega$,
but $n_{(0)}^\mu=(0,1,0,0)$ and
\begin{equation}
 \begin{split}
  T_{(0)}{}^\nu{}_{r\nu} &=
  2/r
  \,.
 \end{split}
\end{equation}
Combining all the terms, the subtraction give the same result as the subtraction of GR
\begin{equation}\label{eq:tegr-subtraction}
 \begin{split}
 -\frac{1}{8\pi} \int_{\partial {\cal M}} \dd^3 y \sqrt{h^{(0)}} \, n^{(0)}_\mu T_{(0)}{}^\mu
 &= \beta \Bigl(-r_0 + \frac{1}{2} r_s\Bigr) + O(r_0^{-1})
 \,.
 \end{split}
\end{equation}
In the end, we have that the Euclidean action evaluated with the Weitzenb\"ok connection gives $S_T [e] = \beta \frac{r_s}{4}$ just like in GR. Needless to say, the same steps of the previous sections can be followed to reproduce the thermodynamical results. These includes also the computation away from equilibrium, which also would give the same result, by a slight generalization of the above presentation which takes into account that the volumetric integral of $\mathring{R}$ is sensitive to the conical singularity. Notice also that a different choice of frame might have lead to different subtraction \cite{Fiorini:2023axr}.

So far we have just followed the usual procedure for subtracting divergences as we do in GR \cite{Oshita:2017nhn,BeltranJimenez:2018vdo}. However, unlike GR (where the bulk either vanishes or is determined by the conical singularity), the boundary term in teleparallel gravity equals the bulk integration due to Eq.~\eqref{eq:tegr-teleparallelism}, namely:
\begin{equation}
 \frac{1}{16\pi}
 \int_{\mathcal M}\mathrm{d}^4x \, \sqrt{-g} \,
 \mathring{\mathbb T}
 =
 -\frac{1}{8\pi} \int_{\partial{\cal M}} \dd^3 y \sqrt{h} \, n_\mu T^\mu
 \ .
\end{equation}
This opens up the possibility of computing the Euclidean action directly from the bulk, without ever referring to surface integrals.
This gives
\begin{equation}\label{eq:Tring-integral}
 \begin{split}
 \frac{1}{16\pi} \int_{\cal M} \dd^4 x \sqrt{g} \, \mathring{\mathbb{T}}
 &= \beta \Bigl(-\frac{1}{4}r_0 + \frac{1}{4} r_s\Bigr) 
 \,,
 \end{split}
\end{equation}
which agrees with the finite part of the difference between \eqref{eq:tegr-boundary} and \eqref{eq:tegr-subtraction}, but also displays a linear divergence for $r_0\to \infty$. The delicate point is whether a subtraction exists that has the form of a bulk integral. Following a prescription by Krssak and Pereira, we show later that this can be done at the price of introducing a reference vierbein everywhere on ${\cal M}$ and choosing a special spin-connection that differs from the Weitzenb\"ock one \cite{Krssak:2015rqa}.

\subsection{Euclidean action in STEGR}\label{sect:quasilocal-STEGR}

A less-known teleparallel equivalent theory of gravity, known as symmetric TEGR (abbreviated STEGR), is the one in which the connection $\nabla$ is kinematically required to be flat, symmetric (thus torsionless),
but not metric-compatible \cite{Nester:1998mp,Mol:2014ooa}. Noncompatibility is quantified by the nonmetricity tensor, which is defined 
\begin{equation}
 \begin{split}
  Q(v)_{\mu\nu} = \nabla_v g_{\mu\nu} \,,
 \end{split}
\end{equation}
and in holonomic components $Q_{\alpha\mu\nu} = \nabla_\alpha g_{\mu\nu}$.
In this case, the requirement of teleparallelism implies
\begin{equation}
 \begin{split}
  0=g^{\beta\nu} R[\nabla]^\alpha{}_{\beta \alpha \nu}
  =  \mathring{R} + \mathring{\mathbb{Q}} + \mathring{\nabla}_\mu (Q^\mu-\tilde{Q}^\mu)
  \,,
 \end{split}
\end{equation}
where we introduced the nonmetricity scalar
\begin{equation}
 \begin{split}
  \mathring{\mathbb{Q}} = \frac{1}{4} Q_{\alpha\mu\nu}Q^{\alpha\mu\nu}
  -\frac{1}{2}Q_{\alpha\mu\nu}Q^{\mu\alpha\nu}
  -\frac{1}{4} Q_\mu Q^\mu +\frac{1}{2} Q_\mu \tilde{Q}^\mu
  \,,
 \end{split}
\end{equation}
and $Q_\mu = Q_{\mu}{}^\alpha{}_{\alpha}$ and $\tilde{Q}_\alpha = Q^\mu{}_{\mu\alpha}$, which are the two independent vector contractions of the nonmetricity itself.
The STEGR action equivalent to GR is then
\begin{equation}
 \begin{split}
  S_{Q} [g,\nabla] = \frac{1}{16\pi} \int_{\cal M} \dd^4 x \sqrt{g} \, \mathring{\mathbb{Q}}
  +S_{\rm sub}
  \,,
 \end{split}
\end{equation}
where, as in TEGR, we introduced a nondynamical subtraction to be determined later.
The action depends in general on the metric $g_{\mu\nu}$ and the connection $\nabla$ through the nonmetricity tensor.

Even though it is less-known than the antisymmetric counterpart, the STEGR formulation can be cast in a form which is older than the Einstein-Hilbert action \eqref{eq:gr-action-euclidean} itself (see the discussion below) by gauge-fixing diffeomorphism invariance.
In fact, the general requirements on the connection can be solved in terms of four functions $\xi^a$ such that the components are
\begin{equation}
 \begin{split}
  \Gamma^\alpha{}_{\beta\gamma} = \frac{\partial x^\alpha}{\partial \xi^a}\partial_\beta\partial_\gamma \xi^a
  \,,
 \end{split}
\end{equation}
where $\frac{\partial x^\alpha}{\partial \xi^a}$ is the inverse matrix of $\frac{\partial \xi^a}{\partial x^\alpha}$.
If we choose the gauge $x^\alpha = \xi^\alpha$, known in the literature as \emph{coincident gauge} \cite{BeltranJimenez:2017tkd}, we have that $\Gamma^\alpha{}_{\beta\gamma}=0$. In this gauge the disformation tensor, defined as the difference $L^\alpha{}_{\beta\gamma}=(\Gamma-\mathring{\Gamma})^\alpha{}_{\beta\gamma}$ becomes the negative of the Levi-Civita connection, $L^\alpha{}_{\beta\gamma}=-\mathring{\Gamma}^\alpha{}_{\beta\gamma}$, or, equivalently, $Q_{\alpha\beta\gamma}=\partial_\alpha g_{\beta\gamma}$.

In practice, in the coincident gauge, the action depends only on the metric, $S_Q[g]$, but it is not covariant, i.e., it is covariant only up to boundary terms as we now briefly discuss. However the equations of motion are covariant, that is, they can be cast in a covariant form. In fact, $S_Q[g]$ is Schr\"odinger's and Einstein's $\Gamma\Gamma$-action \cite{Tomboulis:2017fim} (up to the nondynamical subtraction term) $S_Q[g] = S_{\Gamma\Gamma}[g] + S_{\rm sub}$. The Euclidean $\Gamma\Gamma$-action is defined as\footnote{%
The directional index of the connection is the third one in our notation, so as a $GL(4)$ matrix the components are $\mathring{\Gamma}^\mu{}_{\nu\alpha} = [\mathring{\Gamma}_\alpha]^\mu{}_\nu$. The square brackets imply the commutator of the two matrices.
}
\begin{equation}
 \begin{split}
   S_{\Gamma\Gamma}[g] = \frac{1}{16\pi} \int_{\cal M} \dd^4 x \sqrt{g} \, g^{\nu\rho}[\mathring{\Gamma}_\mu,\mathring{\Gamma}_\nu]^\mu{}_\rho
  \,,
 \end{split}
\end{equation}
and it is equal the Einstein-Hilbert action up to a boundary term $\mathcal{B}$. Consequently, the $\Gamma\Gamma$-action gives Einstein's equations fully expressed in terms of the metric, even though it is explicitly not invariant under general diffeomorphisms. Of course, this invariance can be restored by reintroducing the boundary term $\mathcal{B}$.

In the coincident gauge we can follow similar steps as in the case of TEGR described before.
The requirement of teleparallelism, combined with $\mathring{R}=0$ for an on-shell solution in vacuum and Stokes' theorem, allows us to rewrite the bulk integral of $S_Q$ as the total derivative on the boundary
\begin{equation}
 \begin{split}
  \frac{1}{16\pi} \int_{\cal M} \dd^4 x \sqrt{g} \, \mathring{\mathbb{Q}}
  =
  -\frac{1}{16\pi} \int_{\cal M} \dd^4 x \sqrt{g} \, \mathring{\nabla}_\mu (Q^\mu-\tilde{Q}^\mu)
  = -\frac{1}{16\pi} \int_{\partial{\cal M}} \dd^3 y \sqrt{h} \, n_\mu (Q^\mu-\tilde{Q}^\mu)\,.
 \end{split}
\end{equation}
It is thus natural to make the following choice for the on-shell Euclidean action and its subtraction
\begin{equation}
 \begin{split}
  S_Q[g]
  = -\frac{1}{16\pi} \int_{\partial{\cal M}} \dd^3 y \sqrt{h} \, n_\mu (Q^\mu-\tilde{Q}^\mu)
  +\frac{1}{16\pi} \int_{\partial{\cal M}} \dd^3 y \sqrt{h^{(0)}} \, n^{(0)}_\mu (Q_{(0)}^\mu-\tilde{Q}_{(0)}^\mu)
  \,,
 \end{split}
\end{equation}
which is completely analogous to the manipulation performed for the TEGR case of the previous section.
The computation proceeds similarly to the previous case of TEGR, so we do not repeat it in detail for brevity. Using $Q_{\alpha\beta\gamma}=\partial_\alpha g_{\beta\gamma}$, we can determine
\begin{equation}
 \begin{split}
  (Q-\tilde{Q})_r = f^{-1}f'+4/r
 \end{split}
\end{equation}
which gives the same result as the GHY one in GR when integrated.
The subtraction requires $n_{(0)}^\mu$ and a fixed $f_{0}=f(r_0)$ component as in the TEGR case, and it also gives the same result as the subtraction term of GR when integrated.
The final result is the expected 
$S_Q [g] = \beta \frac{r_s}{4}$, previously obtained both in GR and TEGR, leading to the same thermodynamics.

An important point is that, having fixed diffeomorphism invariance, the bulk integral of $\mathring{\mathbb{Q}}$, that is, the integral of $S_{\Gamma\Gamma}$ itself, depends on the gauge choice. This is, both in principle and in practice, a more severe problem if compared to the TEGR case.
In Schwarzschild coordinates, the volumetric integral is
\begin{equation}\label{eq:ringQ-integral}
 \begin{split}
  \frac{1}{16\pi} \int_{\cal M} \dd^4 x \sqrt{g} \, \mathring{\mathbb{Q}}
  &= -\frac{\beta r_0}{2} + \frac{\beta r_s}{2} 
  \,,
 \end{split}
\end{equation}
and, contrary to the TEGR case, the finite part does not coincide with
the expected result, but, rather, it is twice as much. It is this problematic volumetric integral, rather than the better-behaving boundary one, that has been evaluated in Ref.~\cite{BeltranJimenez:2018vdo}, spawning a subsequent literature to solve the issue. We return to this point in the next section, which is dedicated to bulk integrals.

\section{Bulk integrals and prescriptions}\label{sect:bulk}

In the metric case, the volumetric part of the Euclidean action vanishes for vacuum solutions, thus being controlled solely by the GHY boundary term. One could still construct volumetric definitions of mass and angular momentum via pseudotensors, but at the price of losing covariance.
The evaluation of volumetric integrals thus poses a challenge in the nonlinear regime, especially over the strongly nonlinear regions of black holes, where coordinates might become ill-defined. 
This suggests that the quasilocal approach might be better suited to define the energy-momentum content of gravitational solutions.

In fact, quasilocal/flux relations have the advantage of taking place at the controlled asymptotic regime, where $h_{\mu\nu}$ is infinitesimal.
This is indeed confirmed by our previous computations: when evaluated on-shell, volumetric integrals (over $\mathring{\mathbb{T}}\sim\mathring{\nabla}_\mu T^\mu$ or $\mathring{\mathbb{Q}} \sim \mathring{\nabla}_\mu (Q^\mu-\tilde{Q}^\mu)$) are traded by surface integrals via Stokes' theorem.
The teleparallel approaches thus share the same steps of the method of pseudotensors, and our trust should be put to quasilocal definitions of energy first and foremost.

However, unlike GR (where the bulk integral vanishes), in teleparallel models we have access to the integrals over $\mathring{\mathbb{T}}$ and $\mathring{\mathbb{Q}}$, thus allowing one compute them directly. As noted before, volumetric integrals might be divergent, thus a subtraction procedure is required. Differently from GR, this subtraction may be done locally in (S)TEGR theories, as we shall show later.

\subsection{A working prescription for bulk integrals in TEGR}\label{sect:bulk-TEGR}

In TEGR neither the volumetric integral of $\mathring{\mathbb{T}}$ nor that of $\mathring{\nabla}_\mu T^\mu$ are zero on-shell, so it is natural to ask whether it is possible to estimate the Euclidean action from them.
In the case of TEGR, the result the volumetric integral is given in \eqref{eq:Tring-integral} and clearly requires a subtraction being divergent for $r\to \infty$. Even so, it is not clear how to recover from Eq.~\eqref{eq:Tring-integral} the standard result that we previously obtained from the boundary terms.

An argument by Krssak and Pereira becomes useful for this purpose \cite{Krssak:2015rqa}.
The connection of TEGR represents a purely inertial effect, so there are infinite pairs of solutions $(e^a{}_\mu, \omega^a{}_{b\nu})$ to the field equations, corresponding to local transformations of the connection 
$\omega^a{}_{b\nu}$, which are fixed in the pure vierbein formalism when choosing the Weitzenb\"ock connection.
It is possible to separate inertial effects from gravitational effects, locally, by appropriately choosing the spin connection. The first step is to choose a reference vierbein $\overline{e}^a{}_\mu$. The latter could be understood as the limit of the would-be solution in absence of gravitational interactions, i.e., $G\to 0$, where $G$ is Newton's constant, which has the expected asymptotic behavior in asymptotically flat solutions.\footnote{%
Notice that we have been working in units in which $G=1$, so for taking the limit it is first necessary to restore it in the Schwarzschild solution by replacing $f(r)=1-2 m G/r$.}
Then, we require that the torsion tensor $T[\overline{e},\omega] \equiv D[\omega] \overline{e}$ of the pair $(\overline{e}^a{}_\mu, \omega^a{}_{b\nu})$ vanishes,
which equivalently implies that
\begin{equation}
 \begin{split}
  \omega^a{}_{b\nu} \equiv \mathring{\omega}^a{}_{b\nu}[\overline{e}^a{}_\mu]
  \,,
 \end{split}
\end{equation}
where $\mathring{\omega}^a{}_{b\nu}[\overline{e}^a{}_\mu]$ is the spin-connection of the reference vierbein. Using this connection, instead of the Weitzenb\"ock choice, we have that the torsion of the pair $(e^a{}_\mu, \omega^a{}_{b\nu})$, becomes
\begin{equation}
 \begin{split}
  T^a{}_{bc} = \omega^a{}_{cb}-\omega^a{}_{bc} + E^\mu{}_b E^\nu{}_c (\partial_\mu e^a{}_\nu-\partial_\nu e^a{}_\mu)
 \end{split}
\end{equation}
which differs from the Weitzenb\"ock choice because of the first two terms.
In practice, the asymptotic frame $\overline{e}^a{}_\mu$ introduces a spin-connection, relative to which we define the actual noninertial torsion.
The claim is that the TEGR action
\begin{equation}
 \begin{split}
  S'_T[e;\overline{e}] &= \frac{1}{16\pi} \int_{\cal M} \dd^4 x \sqrt{g} \, \mathring{\mathbb{T}}
  \,,
 \end{split}
\end{equation}
gives finite results without the need of subtractions. This is shown explicitly for the ADM mass and angular momentum in Ref.~\cite{Krssak:2015rqa}. Notice that $S'_T[e;\overline{e}]$ is essentially a bi-field
functional and the role of the reference vierbein can be assimilated to that of the metric $\eta_{\mu\nu}$ when defining pseudotensors.

In the practical case of Schwarzschild, we formally have that $ \lim_{G\to0} f(r)=1$. Thus, the reference vierbein is chosen to be $\overline{e}^a{}_\mu={\rm diag}(1,1,r,r \sin{\theta})$, which defines the purely inertial spin-connection, and literally gives the metric $\eta_{\mu\nu}$ used to define $h_{\mu\nu}$. The relation allows us to determine the torsion in terms of the true vierbein and the spin-connection of the reference one.
Using the newly determined torsion, we find directly
\begin{equation}
 \begin{split}
  \frac{1}{16\pi} \int_{\cal M} \dd^4 x \sqrt{g} \, \mathring{\mathbb{T}}
  &= \frac{r_s \beta}{4} = \frac{\beta^2}{16\pi}
 \end{split}
\end{equation}
from which we can derive the usual thermodynamical properties without having to perform any further subtraction. It interesting to see how this is reflected in the integral of the total divergence, which, by construction, must give the same result. We have that there are \emph{two} boundaries, at $r=r_s$ and at large $r$
\begin{equation}
 \begin{split}
  \frac{1}{16\pi} \int_{\cal M} \dd^4 x \sqrt{g} \, \mathring{\mathbb{T}}
  &=
  -\frac{1}{8\pi} \int_{r} \dd^3 y \sqrt{h} \, n_\mu T^\mu
  -\frac{1}{8\pi} \int_{r_s} \dd^3 y \sqrt{h} \, n_\mu T^\mu
 \end{split}
\end{equation}
and they contribute equally to the final result, i.e.,
\begin{equation}
 \begin{split}
 -\frac{1}{8\pi} \int_{r} \dd^3 y \sqrt{h} \, n_\mu T^\mu
  =-\frac{1}{8\pi} \int_{r_s} \dd^3 y \sqrt{h} \, n_\mu T^\mu
  = \frac{r_s \beta}{8}
  \,.
 \end{split}
\end{equation}
Notice that the normal vector $n^\mu$ along the $r$ direction changes sign on the two surfaces so that it is outward pointing, i.e., $n^\mu \partial_\mu \sim \partial_r$ at $r\to \infty$ and $n^\mu \partial_\mu \sim -\partial_r$ at $r=r_s$.

One may wonder why is there an inner boundary even when $\beta=\beta_H$, in which case there is no conical singularity. Our interpretation is that this happens because we have introduced the reference vierbein $\overline{e}^a{}_\mu$
for computing the inertial connection, and the reference vierbein does have a boundary at $r=r_s$ in the Euclidean patch. This inner boundary is the one which is taken into account by the integral (given that $r>r_s$, $\overline{e}^a{}_\mu$ is the vierbein of a portion of flat space with a removed inner sphere). In this way Stokes' theorem works properly, without having to rely only on the quasilocal relation. 
However, we do not have an obvious explanation for why the two surfaces should contribute equally.

Before proceeding to the analysis of volumetric integrals of STEGR, we briefly use this regularization procedure to discuss a connection with the Landau-Lifshitz energy momentum pseudotensor.

\subsection{A digression on the Landau-Lifshitz pseudotensor in TEGR}

In order to establish an analogy between the Landau-Lifshitz pseudotensor and a similar tensor appearing in TEGR, we need to recall first the construction of the former.
For this section we return temporarily to Lorentzian signature.
The key equation in the construction of the Landau-Lifshitz pseudotensor $t^{\mu\nu}$ comes from rewriting Einstein's equations as
\begin{equation}\label{gen_en_mom_tens}
 \partial_\rho h^{\mu\nu\rho}=(-g) (T^{\mu\nu} + t^{\mu\nu})
 \,,
\end{equation} 
where we have included the matter energy-momentum tensor $T^{\mu\nu}=\frac{2}{\sqrt{g}} \frac{\delta S_{m}}{\delta g_{\mu\nu}}$ for completeness and introduced the pseudotensor $h^{\mu\nu\rho}$, which is defined as
\begin{equation}\label{eq:h3-def}
 \begin{split}
 h^{\mu\nu\rho} =
 \partial_\gamma \lambda^{\mu\nu\rho\gamma}
 \, , \qquad
 \lambda^{\mu\nu\rho\gamma} = \frac{1}{16\pi }(-g) (g^{\mu\nu} g^{\rho\gamma} 
 -g^{\mu\rho}g^{\nu\gamma} ) \, .
 \end{split}
\end{equation}
In practice, the symmetric pseudotensor $t^{\mu\nu}$ is defined implicitly by inserting \eqref{eq:h3-def} in \eqref{gen_en_mom_tens} and comparing with Einstein's equations, resulting in a noncovariant expression that depends on contractions of the components of the Levi-Civita connection. The explicit form of $t^{\mu\nu}$ is very complicate, but, fortunately, not needed in practical computations.
The important property is that $h^{\mu\nu\rho}= - h^{\mu\rho\nu}$, implying that we can define a conserved current 
\begin{equation}\label{LLempt_cons_current}
 \partial_\nu \bigl[(-g) (T^{\mu\nu} + t^{\mu\nu})\bigr]
 =
 \partial_\nu\partial_\rho h^{\mu\nu\rho}=0
 \, ,
\end{equation}
and clarifying the role of the combination $(-g) (T^{\mu\nu} + t^{\mu\nu})$
as the one producing a true conservation law.
The charges combine as the four momentum in some asymptotic frame, defined as
the integral over spacelike hypersurfaces
\begin{equation}\label{LLempt_cons_charges}
 P^{\mu}
 \equiv
 \int \bigl[(-g) (T^{\mu0} + t^{\mu0})\bigr] \dd^3 x\,,
\end{equation}
which is conserved and can be written
as a surface integral using the equations of motion. We have numerical labels for the component to avoid unnecessary confusion in the notation. Introducing
a sphere at large $r$ with metric $\sigma_{AB}$ and coordinates $\theta^A=(\theta,\varphi)$, we have
\begin{equation}\label{LLempt_cons_charges-2}
 P^{\mu}
 \equiv
 \int \partial_\rho h^{\mu0\rho} \dd^3 x
 = \int h^{\mu0i} n_i \sqrt{\sigma} \dd ^2 \theta
 \,.
\end{equation}
The zeroth component coincides with \eqref{eq:weinberg-energy} asymptotically after substituting the explicit form of the pseudotensor $h_{\mu\nu\rho}$.

Now we want to show that, using the field equations of TEGR, it is possible to get conserved currents as well in complete analogy with the above procedure. Let $\Theta_{a}{}^{\rho}$ be the generalized matter's energy-momentum tensor, defined $\Theta_{a}{}^{\rho} \equiv \frac{1}{\underline{e}}\frac{\delta S_{\rm mat}}{\delta e^a{}_\rho}$ for $\underline{e}=\det(e^a{}_\rho)$. The equations of motion of TEGR can be written as
\begin{equation}
\partial_{\sigma}\left( \underline{e} S_{a}{}^{\rho\sigma} \right)
=
8 \pi e(\Theta_{a}{}^{\rho} - J_{a}{}^{\rho})
\, ,
\end{equation}
where $S_{a}{}^{\rho\sigma}$ is the superpotential and the structure should be compared to \eqref{gen_en_mom_tens}. In a holonomic basis, $S_{a}{}^{\rho\sigma}=e^{a}{}_{\rho}S^{\rho\mu\nu}$, the superpotential depends on torsion and contortion
\begin{equation}\label{superpotential}
S^{\rho\mu\nu}=K^{\mu\nu \rho}-g^{\rho\nu}T^{\sigma\mu}{}_{\sigma}+g^{\rho\mu}T^{\sigma\nu}{}_{\sigma}\,.
\end{equation}
Importantly, we have also introduced the current $J$, which is defined as
\begin{equation}\label{J_current}
\underline{e} J_{a}{}^{\mu}=  \frac{\partial \mathcal{L}_T}{\partial e^{a}{}_{\mu}} \, ,
\end{equation} 
where $\mathcal{L}_T=\frac{\underline{e}}{16\pi}\mathring{\mathbb{T}}$ is the TEGR Lagrangian. It should be clear that the current plays the same role as $\tau^{\mu\nu}$ in the Landau-Lifshitz construction.

Now we can clarify the analogy between TEGR and the Landau-Lifshitz construction. For simplicity we consider the pure-gravity case in which $\Theta=0$.
Thanks to the antisymmetry in $S_{a}{}^{\rho\sigma}=-S_{a}{}^{\sigma\rho}$ of the superpotential, we can see that $\underline{e}J_{a}{}^{\rho}$
is a conserved current, i.e.,
\begin{equation}\label{TEGR_cons_current}
\partial_{\rho}(\underline{e} J_{a}{}^{\rho})=-\frac{1}{8\pi}\partial_{\rho}\partial_{\sigma}\left(\underline{e} S_{a}{}^{\rho\sigma} \right)=0 \, .
\end{equation}
Comparing the conservation laws \eqref{LLempt_cons_current} and \eqref{TEGR_cons_current}, we confirm the analogy between $J$ and the pseudotensor $t$. Similarly, we can identify the three-indexed tensors 
$
-\frac{\underline{e}}{8\pi} S^{a\rho\sigma} \sim h^{\mu\nu\rho}.
$
Likewise Eq.~\eqref{LLempt_cons_charges}, we define the conserved charge contained in a sphere as a Lorentz vector in an asymptotic frame. Using the same manipulations as the Landau-Lifshitz case we have
\begin{equation}\label{TEGR_gravitational_empt}
 P^{a}
 =
 -\frac{1}{8\pi}\oint_{S_{2}} S^{a0i}  n_{i} \sqrt{\sigma} \dd^2\theta 
\, ,
\end{equation}
for which Stokes' theorem has been used and the integration is performed as usual on a boundary $2$-sphere at large constant $r=r_0$ embedded in a spacelike hypersurface at constant $t$. The component $a=0$ should be interpreted as the gravitational energy.

Let us compute the energy using Eq.~\eqref{TEGR_gravitational_empt} and coordinates $x^\mu=(t,r,\theta,\varphi)$ and the Weitzenb\"ock connection. The normal vector has only the $r$-component, $n_1=f^{-1/2}$. The only surviving component $S^{00i}$ of the superpotential in the contraction for the Schwarzschild solution is
\begin{equation}
S^{001}
=
K^{010}+g^{00}T^{\sigma 1}{}_{\sigma}
=
-f^{-1}\frac{r_s}{r^{2}}
\end{equation}
because the contortion component is related to the torsion as $K^{010}=T^{001}$.
Contracting $S^{\rho\mu\nu}$ with $e^{a}{}_{\rho}$, we obtain the expected energy
\begin{equation}
 \begin{split}
 P^{\hat{0}}&=-\frac{1}{8\pi}\oint_{S_{2}}    \Big(-f^{-\frac{1}{2}}\frac{r_s}{r^{2}}\Big)  f^{-\frac{1}{2}} r^{2}\sin\theta \dd\theta \dd\varphi
= m + {\cal O}(r_0^{-1})
 \end{split}
\end{equation}
where the hat stands for $ P^{\hat{0}}=P^{a=0}$, denoted differently to stress that it is a Lorentz index rather than a holonomic one. As expected $m$ is the energy $E$ of the Schwarzschild black hole
and the two frames are identified asymptotically if $e^a{}_\mu \to \delta^a{}_\mu$ at large $r_0$. Since we arrived at this relation in the Lorentzian signature, it is necessary to integrate $T_H \dd S=\dd E$ using Hawking's temperature $T_H= 1/\beta_H$ to correctly obtain the Bekenstein-Hawking entropy $S=\pi r_s^2=\frac{A}{4}$.

\subsection{Issues with the prescription in the case of STEGR}\label{sect:bulk-STEGR}

Now let us try to frame a possible subtraction that may work for the STEGR case.
As before, the integral of $\mathring{Q}$ is divergent, as given in \eqref{eq:ringQ-integral}.
The \emph{naive} generalization of Krssak-Pereira's prescription
to the case of STEGR would be to introduce a reference metric $\overline{g}_{\mu\nu}$ and choose a symmetric connection $\nabla$ for which the nonmetricity $Q[\overline{g},\nabla]=0$. By construction it must be that
the connection $\nabla$ is the Levi-Civita connection of the reference metric, so $\Gamma^\mu{}_{\nu\rho}=\mathring{\Gamma}[\overline{g}]^\mu{}_{\nu\rho}$.
With this prescription, the nonmetricity of $g_{\mu\nu}$ becomes $Q_{\mu\nu\rho} = \partial_\mu g_{\nu\rho}- \Gamma^\alpha{}_{\mu} g_{\alpha\rho}- \Gamma^\alpha{}_{\mu} g_{\nu\alpha}$. Repeating the integral of $\mathring{Q}$ with the new form of the nonmetricity gives
\begin{equation}
 \begin{split}
  \frac{1}{16\pi} \int_{\cal M} \dd^4 x \sqrt{g} \, \mathring{\mathbb{Q}}
  &= 0 
  \,,
 \end{split}
\end{equation}
which obviously does not give the expected result.
Another subtraction procedure has been discussed in Ref.~\cite{Krssak:2024kva} using the vierbein formalim, and results in a finite part that overestimates the entropy by a factor two, similarly to the bulk integral given in Sect.~\ref{sect:quasilocal-STEGR}.

The problem with estimating the entropy using gauge-fixed STEGR actions is that, by its nature, the gauge-fixing of diffeomorphisms results in coordinate dependent results for the bulk integrals, as noted in Ref.~\cite{BeltranJimenez:2018vdo}. Discussions on this point also appear in Ref.~\cite{BeltranJimenez:2019bnx}, but the final result is ultimately estimated from the total divergence term as we have done in Sect.~\ref{sect:quasilocal-STEGR}.

A correct estimate of the entropy appears in Ref.~\cite{Heisenberg:2022nvs}, in which the authors use Wald's formula by correctly adapting it to the case of STEGR. The methods of Ref.~\cite{Heisenberg:2022nvs}
are quite powerful, but also rather technical, so they depart from the simplicity of just evaluating an Euclidean action over a background.
In the next section we give a simpler prescription which works for the case of Schwarzschild and only requires a small departure from the coincident gauge for the St\"uckelberg fields $\xi^a$.

\subsubsection{Minimization method}

As anticipated, we now present a prescription for obtaining the expected thermodynamics that works, based on the minimization of the St\"uckelberg fields $\xi^a$ appearing in the STEGR connection.
To begin with, recall the general solution to the STEGR connection
$
 \Gamma^\alpha{}_{\beta\gamma} = \frac{\partial x^\alpha}{\partial \xi^a} \partial_\beta \partial_\gamma \xi^a
$
for a set of four functions $\xi^a$ under the assumption that the matrix $\partial_\alpha \xi^a$ is invertible.
The $\xi^a$ can be interpreted as a set of St\"uckelberg fields to implement diffeomorphism invariance.

We covariantize the expression of the STEGR connection using the Levi-Civita connection and treating $\xi^a$ as scalar functions
\begin{equation}
 \Gamma^\alpha{}_{\beta\gamma} = \frac{\partial x^\alpha}{\partial \xi^a} \mathring{\nabla}_\beta \partial_\gamma \xi^a + \mathring{\Gamma}^\alpha{}_{\beta\gamma}\,,
\end{equation}
which is motivated by the logic that the coordinates are locally scalar functions.
From the above expression, we deduce the disformation tensor
\begin{equation}
 L^\alpha{}_{\beta\gamma} \equiv \Gamma^\alpha{}_{\beta\gamma} -\mathring{\Gamma}^\alpha{}_{\beta\gamma} = \frac{\partial x^\alpha}{\partial \xi^a} \mathring{\nabla}_\beta \partial_\gamma \xi^a
 \,,
\end{equation}
then the STEGR action can be regarded as a functional of either the nonmetricity tensor or the disformation tensor.
In fact, STEGR action can be written conveniently in terms of the disformation
\begin{equation}
 \begin{split}
  S_{Q} [g,\xi]
  &=
  \frac{1}{16\pi} \int_{\cal M} \dd^4 x \sqrt{g} \, g^{\nu\rho}[L_\mu,L_\nu]^\mu{}_\rho
  +S_{\rm sub}
  \,,
 \end{split}
\end{equation}
where $[L_\mu,L_\nu]^\mu{}_\rho = L^\mu{}_{\alpha\mu}L^\alpha{}_{\rho\nu}-L^\mu{}_{\alpha\nu}L^\alpha{}_{\rho\mu}$ and $S_{\rm sub}$ is yet to be specified. The action depends on the metric $g_{\mu\nu}$ and on the functions $\xi^a$. It is still true that the first term becomes Einstein-Schr\"odinger's $\Gamma\Gamma$-action in the gauge $\xi^\alpha=x^\alpha$.

There is an enormous amount of freedom in the possible choices for the functions $\xi^a(x)$ and we have experimented with some possible options.
In the following we present a choice that we believe is physically motivated and not based on cherry-picking.
To evaluate the volumetric integral, we choose to not fix the gauge entirely, but rather leave a functional parametric dependence. In Schwarzschild coordinates we choose
\begin{equation}\label{eq:noncoincident-coordinates}
 \xi^0 = t(1+h(r))\,, \qquad \xi^1=r\,,\qquad \xi^2 = \theta\,, \qquad \xi^3 = \varphi\,,
\end{equation}
which depend only on an undetermined scalar function $h(r)$ with radial symmetry. If $h(r)=0$, then the gauge becomes the coincident gauge as it implies $\xi^\alpha=x^\alpha$, but we \emph{only} require $h(r_s)=0$ so that the coordinates correctly capture the near-horizon Schwarzschild ones in the limit $r\to r_s$. The parametrization is motivated by symmetry and by the fact that
the standard result captures the ADM mass with a multiplicative factor, so the zeroth component of the four-momentum is incorrect.

We now evaluate the angular integrals of the first term
using $g_{\mu\nu} = {\rm diag}(f,f^{-1},r^2,r^2 \sin \theta)$
for general $h(r)$ as
\begin{equation}
 \begin{split}
 \frac{1}{16\pi} \int_{\cal M} \dd^4 x \sqrt{g} \, g^{\nu\rho}[L_\mu,L_\nu]^\mu{}_\rho
 &= -\int_{r_s}^{\infty} {\cal L}_h(h,h';f) \, \dd r
 \,.
 \end{split}
\end{equation}
The right hand side is the integral over the radial coordinate of a function of $h(r)$ and $h'(r)= \frac{\partial h}{\partial r}$, which depends parametrically on $f(r)$.
We interpret the right hand side as an action for the function $h(r)$,
so that ${\cal L}_h$ is its Lagrangian.
The explicit form is
\begin{equation}
 \begin{split}
 {\cal L}_h(h,h';f) = \frac{\beta r^2}{4} \frac{f(r)}{(1+h(r))^2} h'(r)^2
 -\frac{\beta r}{4} \frac{2 f(r)+r f'(r)}{(1+h(r))^2} h'(r)
 +\frac{\beta}{2}(f(r)+r f'(r))\,,
 \end{split}
\end{equation}
and the last term is the one responsible for overestimating the logarithm of the partition function.

Minimization of the action with respect to $h(r)$ is then akin to minimization with respect to the parameter $m$ when going to equilibrium as done in Sect.~\ref{sect:thermodynamics}. Using the Euler-Lagrange equations, we require
\begin{equation}
 \begin{split}
 \frac{\delta}{ \delta h} \int {\cal L}_h(h,h';f)\dd r=0
 \qquad \Longrightarrow \qquad \frac{\partial}{\partial h}{\cal L}_h(h,h';f) - \frac{\dd}{\dd r}\frac{\partial}{\partial h'}{\cal L}_h(h,h';f)=0
 \,,
 \end{split}
\end{equation}
which gives a second order differential equation in $h(r)$. We choose the boundary conditions $h(r_s)=0$ and $r h'(r)|_{r_s}=1$, for which, when $f(r)=1-r_s/r$ the Euler-Lagrange quations have solution
\begin{equation}
 \begin{split}
 h(r) = \frac{r-r_s}{r_s}
 \,,
 \end{split}
\end{equation}
but we have checked that the final result does not actually depend on the boundary condition on $h'(r)$.
We now evaluate the action on-shell for this solution using a cutoff at large-$r$ and find 
\begin{equation}\label{eq:lagrangian-stegr-onshell}
 \begin{split}
 -\int_{r_s}^{r_0} {\cal L}_h(h,h';f) \, \dd r
 &= \frac{\beta}{4}(-r_0+r_s) +{\cal O}(r_0^{-1})
 \,.
 \end{split}
\end{equation}
We stress once more that the result does not depend on the boundary condition on $h'(r)$, but only on the choice $h(r_s)=0$. The finite part of the result is $\frac{\beta r_s}{4}$, which leads to the standard thermodynamics should we be able to subtract the divergent part.

For the subtraction term we repeat the same procedure, but we evaluate everything in flat space instead, i.e., for $f(r)=1$. In this case, using the boundary conditions $h(r_s)=h'(r_s)=0$, we find the solution
\begin{equation}
 \begin{split}
 h(r) = \frac{r \, {\rm e}^{\frac{r_s-r}{r}}-r_s}{r_s}\,,
 \end{split}
\end{equation}
and the on-shell action becomes
\begin{equation}\label{eq:lagrangian-stegr-onshell-sub}
 \begin{split}
 S_{\rm sub} \equiv \left.\int_{r_s}^{r_0} {\cal L}_h(h,h';f) \, \dd r \right|_{f=1}
 &= \frac{\beta r_0}{4} +{\cal O}(r_0^{-1})
 \,.
 \end{split}
\end{equation}
Combining the contributions \eqref{eq:lagrangian-stegr-onshell} and \eqref{eq:lagrangian-stegr-onshell-sub} we find a result without divergence in the limit $r_0\to \infty$
\begin{equation}
 \begin{split}
  S_{Q} [g,\xi] &= \frac{\beta r_s}{4}
  \,,
 \end{split}
\end{equation}
as expected for obtaining the standard thermodynamics. We have checked that modifications of the above procedure including a radial function, e.g., modifying $\xi^1 = r(1+ \tilde{h}(r))$ instead of $\xi^0$
in \eqref{eq:noncoincident-coordinates} do not change the standard result.
We only have an incomplete understanding of the above procedure, therefore, even though it provides us with the expected answer, we believe that it should be regarded only as a qualitative approach to estimate the partition-function. A formally sound approach based on Wald's formula that works for STEGR in the coincident gauge has been developed by Heisenberg et al.\ in Ref.~\cite{Heisenberg:2022nvs} and gives the expected free energy.

\section{Corrections to the Schwarzschild entropy in New GR}\label{sect:corrections}

New GR is a modification of the TEGR model which is allowed by modifying the relative coefficients of the scalar $\mathring{\mathbb{T}}$ \cite{Hayashi:1979qx}. One of the main original motivations for exploring these theories stems from M\o ller's attempt to construct a gravitational energy-momentum tensor that yields an energy density invariant under purely spatial coordinate transformations, while ensuring that the total energy-momentum transforms as a four-vector under the Lorentz group \cite{Moller:1961jj}. This idea was later developed further by Pellegrini and Pleba\'{n}ski, who first constructed the most general Lagrangian for these theories \cite{Pellegrini_Plebansky}.
In this work, we consider New GR as a modification of TEGR in the same spirit as as one often considers modified theories derived from metric-based GR.

The complete New GR action is defined as
\begin{equation}
 \begin{split}
  S_{NGR} [e,\omega] = \frac{1}{16\pi} \int_{\cal M} \dd^4 x \sqrt{g} \, {\mathbb{T}}
 \end{split}
\end{equation}
where the scalar
\begin{equation}
 \begin{split}
  {\mathbb{T}} = -\frac{c_1}{4} T_{\alpha\mu\nu}T^{\alpha\mu\nu}
  -\frac{c_2}{2}T_{\alpha\mu\nu}T^{\mu\alpha\nu}
  + c_3 T_\mu T^\mu
 \end{split}
\end{equation}
is chosen such that $\mathbb{T}|_{c_i=1}= \mathring{\mathbb{T}}$, so $c_i=1$ becomes the (equivalent to) GR limit. In general, the New GR model propagates more degrees of freedom if compared to GR, including parts of a Kalb-Ramond like field. In the physical limit, discussed by van Nieuwenhuizen in \cite{VanNieuwenhuizen:1973fi}, $(2 c_3-c_1-c_2)=0$, we have that symmetric and antisymmetric parts of the linearized vierbein are decoupled, so the quadratic part of the action is essentially the sum of a massless Pauli-Fierz field (which includes the graviton as $2^+$ component) and a Kalb-Ramond-like field.
Here ``physical'' is used in the sense that the given limit
is believed to be the only one consistent with quantization of the gravitational fields in that, otherwise, the theory would be plagued by kinematical ghosts \cite{VanNieuwenhuizen:1973fi}.

The linearized vierbein can be parametrized as $2E_{\nu a}\delta e^{a}{}_\mu = h_{\mu\nu}+b_{\mu\nu}$,
where $h_{\mu\nu}$ is the symmetric part and $b_{\mu\nu}$ is the Kalb-Ramond field. Importantly, in the physical limit, linearized diffeomorphisms are then enhanced to two separate symmetries for either field, i.e., $h_{\mu\nu} \to h_{\mu\nu}+2\partial_{(\mu} \xi_{\nu)}$ and $b_{\mu\nu} \to b_{\mu\nu}+2\partial_{[\mu} \zeta_{\nu]}$ hold separately.
Whether this enhanced symmetry comes from the linearization of a nonlinear enhancement of the diffeomorphisms group is not clear.

The Newtonian potential can be obtained directly from the symmetric part of the equations of motion of the New GR action coupled to a standard energy-momentum tensor, e.g., in a (Lorentzian) multipole expansion.
A back-of-the-envelope computation in the linearized limit reveals
\begin{equation}
 \begin{split}
 \Phi(t,r) \sim \frac{4 c_3-c_1-c_2}{8\pi} \frac{1}{r} \int \dd^3x \, \rho(t,\vec{x}) = \frac{M}{4\pi r}
 \,,
 \end{split}
\end{equation}
where $\Phi$ is the gauge-invariant Bardeen potential $\Phi(t,r) \sim h_{00}$, and $\rho =T_{00}$ is a source concentrated in some region of space observed at a large distance $r$ \cite{Bardeen:1980kt}.
We have the obvious relation between the leading term of the expansion and the integral of the density source, $M = (4 c_3-c_1-c_2)/2 \int \dd^3x \, \rho(t,\vec{x})$. Putting together the Newtonian analysis and the van Nieuwenhuizen limit, it is natural to reparametrize the couplings $c_i$ as
\begin{equation}\label{eq:reparametrization-newgr-couplings}
 \begin{split}
 c_1= \lambda\,, \qquad c_2= 2-\lambda\,, \qquad c_3=1\,,
 \end{split}
\end{equation}
for some leftover coupling $\lambda$, which we will ultimately use at the end of computations. The above limit ensures $M= m \equiv \int \dd^3x \, \rho $ and the absence of ghosts.
Some spherically symmetric solutions of New GR in this limit have been classified by
Asuk\"ula et al.\ in Ref.~\cite{Asukula:2023akj} and include the Schwarzschild solution as well as two more complicate classes of solutions. We now concentrate on the Schwarzschild case, although it might be interesting to apply the following to the other classes.

We choose to work directly in the Krssak-Pereira prescription of Sect.~\ref{sect:bulk-TEGR}, so we assume a reference frame $\overline{e}^a{}_\mu$, either determined a posteriori as the noninteracting limit $G=0$, or as the putative asymptotically flat frame. Furthermore, we choose the spin-connection to be the one induced by the reference frame, that is,
\begin{equation}
 \begin{split}
  T^a{}_{bc} &= \omega^a{}_{cb}-\omega^a{}_{bc} + E^\mu{}_b E^\nu{}_c (\partial_\mu e^a{}_\nu-\partial_\nu e^a{}_\mu)
  \,,
  \\
  \omega^a{}_{b\nu} &= \mathring{\omega}^a{}_{b\nu}[\overline{e}^a{}_\mu]
  \,.
 \end{split}
\end{equation}
The limit $G\to 0$ is potentially more tricky in this case, since there are more couplings in the New GR action, but we still take $\overline{e}^a{}_\mu={\rm diag}(1,1,r,r \sin{\theta})$ from the asymptotic region.
As for the vierbein itself, on the Euclidean patch of Schwarzschild we take $e^a{}_\mu={\rm diag}(f^{1/2},f^{-1/2},r,r \sin{\theta})$.
The vierbeins are diagonal thanks to the fact that we are working with the Schwarzschild solution, while the other cases would feature more compplicate structures.

The result is that the integral of $S_{NGR}$ is divergent in proximity of the horizon for general $c_i$.
For this reason, we introduce a dimensionless cutoff by replacing
$r_s \to r_s(1+\epsilon)$. In the limit $\epsilon\ll 1$, the regulated Euclidean action becomes
\begin{equation}
 \begin{split}
  \frac{1}{16\pi} \int_{r_s(1+\epsilon)}^{\infty}\dd r \int_0^\pi \dd\theta \int_0^{2\pi} \dd\varphi  \sqrt{g} \, {\mathbb{T}}
  = 
  (4 c_3-c_1-c_2)\frac{r_s \beta}{8} + (2 c_3-c_1-c_2)\frac{r_s \beta}{64} \log(\epsilon')
  +{\cal O}(\epsilon')
 \end{split}
\end{equation}
where we also rescaled the cutoff as $\epsilon'= 2^{-16}\epsilon$ for convenience.
Having introduced a cutoff, we use the standard terms ``renormalized'' and ``bare'' of QFT, although loosely in this context. Similar logarithmic corrections are familiar when integrating-out quantum fluctuations \cite{Fursaev:1994te}.

From the partition function, we can define the total renormalized energy
\begin{equation}
 \begin{split}
 M_{\rm ren} = -\frac{\partial}{\partial\beta} \log Z
 = \frac{1}{2} (4 c_3-c_1-c_2) m + \frac{1}{16} (2 c_3-c_1-c_2) \log(\epsilon')
 \,,
 \end{split}
\end{equation}
which differs from the bare mass $m$ that we injected through the solution. However, the leading term of $M_{\rm ren}$, defined $M=\frac{1}{2} (4 c_3-c_1-c_2) m$, is the same mass that appears
in the Newtonian potential that can be obtained directly from the equations of motion of the New GR in the (Lorentzian) multipole expansion,
which is a nice consistency check.
As for the entropy, we find
\begin{equation}
 \begin{split}
 S_{\rm ren} = \Bigl(1-\beta \frac{\partial}{\partial\beta}\Bigr) \log Z
 = \frac{A}{8} (4 c_3-c_1-c_2) + \frac{A}{64} (2 c_3-c_1-c_2) \log(\epsilon')
 \,,
 \end{split}
\end{equation}
expressed in terms of area of the horizon $A=4\pi r_s^2$.

It is tempting to interpret the $\log(\epsilon')$ corrections as caused by a nontrivial renormalization group-like running of the linear combination $2 c_3-c_1-c_2$, although this is a bit speculative. The correction thus appears to be entirely driven by the kinematical ghosts.
More importantly, in the physical limit
in which symmetric and antisymmetric linearized fluctuations of the vierbein decouple, i.e., $2 c_3-c_1-c_2=0$, we see that the divergent parts disappear. Given that the limit $2 c_3-c_1-c_2=0$ has an enhanced symmetry, at least at the linearized level, one may argue that it should be a fixed point of some presumed RG flow, and hence represents a viable UV/IR theory, in which case the divergences of the physical quantities would not present. The final result is that, in the parametrization \eqref{eq:reparametrization-newgr-couplings}, the Bekenstein-Hawking relation $S=A/4$ is preserved with energy $E=M$ for Schwarzschild black holes also in modified in New GR theories.

\section{Conclusions}\label{sect:conclusions}

We have evaluated entropy and energy of static spherically symmetric (Schwarzschild) black holes in various ways using the Euclidean action for the two most important teleparallel theories equivalent to GR.
Our main conclusion is that both the symmetric and the antisymmetric
teleparallel theories lead to the same results as GR, i.e., the Bekenstein-Hawking area law for the entropy and the ADM mass for the energy, iff the same set of basic assumptions are considered. These are the fact that the integrals are evaluated on-shell and that volumetric integrals must be recasted as boundary ones by applying Stokes' theorem.
These two main assumptions ensure that boundary terms of teleparallel theories precisely match those coming from the Gibbons-Hawking-York term of GR (even though teleparallel equivalent theories do not need ``true'' Gibbons-Hawking-York-like terms under Dirichlet boundary conditions).

Our discussion has also prompted us to try and estimate volumetric/bulk integrals of the teleparallel models, because, differently from GR, they are nonzero even on-shell. Using bulk integrals the results are mixed: for the case of the antisymmetric teleparallel theory there exists a  subtraction prescription by Krssak and Pereira that results in a regularized free energy in complete agreement with GR. Instead, for the symmetric theory, the subtraction prescription is less promising, in that it estimated the free energy to either zero or twice the expected value, however we came up with a simple method that gives the expected result when minimizing at equilibrium also some component of a St\"uckelberg field for diffeomorphisms symmetry. Even though the method is surely less formal than other approaches based on Wald's entropy formula, we believe that our approach may be interesting if applied to other solutions given its simplicity, but it first requires a more quantitative physical understanding.

Given that the subtraction procedure works very well for the case of the antisymmetric model, we have applied it also to one extension known as New GR, which has the Schwarzschild black hole as solution in a physically interesting region of its parameter space. In this case, we have observed
that a kinematical ghost instability may drive the free energy to a divergent result, but, if ghosts are kept under control in the parameter space, the final result is that Schwarzschild black holes have the same thermodynamics as in the standard GR case. That said, the space of spherically symmetric solutions of New GR is not limited to Schwarzschild-like black holes, so in the future it would be interesting to compute the entropy of solutions that we have not considered in this work.

\smallskip

\paragraph*{Acknowledgments.}
O.Z.~is grateful to M.~Kr\v{s}\v{s}\'ak for a brief communication that sparked further development of this project.  I.K.~thanks the National Council for Scientific and Technological Development -- CNPq (grant numbers 303283/2022-0 and 401567/2023-0) for financial support.

\appendix

\section{Euclidean Schwarzschild and conical singularity}\label{appendix:euclidean-schwarzschild}

The (Lorentzian) Schwarzschild metric in the coordinates $(t,r,\theta,\varphi)$ is
\begin{equation}
 \begin{split}
  \dd s^2 = -f(r) \dd t^2 + f(r)^{-1} \dd r^2 + r^2 \dd \Omega^2
  \,,
 \end{split}
\end{equation}
with $f(r)=1-\frac{r_s}{r}$ and the Schwarzschild radius is $r_s=2m$. Defining the Euclidean time $\tau= i t$, we find the Euclidean metric
\begin{equation}
 \begin{split}
  \dd s^2|_E = f(r) \dd \tau^2 + f(r)^{-1} \dd r^2 + r^2 \dd \Omega^2\,.
 \end{split}
\end{equation}
Now we prove that the Euclidean section only covers the region $r>2m$,
and that the Euclidean time should be periodic, $\tau\sim\tau+8\pi m$,
in order to avoid a conical singularity at $r=2m$. If the singularity is
avoided then the section is smooth everywhere in the region $r\geq 2m$.

To see that the Euclidean metric covers $r>2m$, consider the standard Kruskal's null coordinates $(U,V)$ and define $T=(U+V)/2$ and $R=(V-U)/2$,
which are well-defined at the horizon. We can Wick rotate $T_E = i T$ and use the expression of $R^2-T^2$ in terms of $r$ to see that
\begin{equation}
 \begin{split}
  R^2+T_E^2 = -f(r) \exp(r/2m)\,.
 \end{split}
\end{equation}
The requirement that $R^2+T_E^2>0$ implies $f(r)<1$, and thus $r>2m$ (inside the horizon one should Wick rotate the coordinate $R$ to obtain an Euclidean metric).

Now concentrate on the region $r \approx 2m$. Consider a new radial coordinate defined $\rho = 4m f(r)^{1/2}$, from which we see
$
 f(r)^{-1/2} \dd r = (r^2/4m^2)\dd \rho
$.
The metric becomes
\begin{equation}
 \begin{split}
  \dd s^2|_E = (\rho/4m)^2 \dd \tau^2 + (r^2/4m^2)^2 \dd \rho^2 + r^2 \dd \Omega^2\,,
 \end{split}
\end{equation}
where $r=r(\rho)$ is determined implicitly from inverting the definition of $\rho=\rho(r)$.
Close to the Schwarzschild radius, $r \approx 2m$, we have
\begin{equation}
 \begin{split}
  \dd s^2|_E \approx (\rho/4m)^2 \dd \tau^2 + \dd \rho^2 + r^2 \dd \Omega^2\,.
 \end{split}
\end{equation}
The first two terms are understood as the metric of an Euclidean flat $2d$ space in polar coordinates, where $\rho$ is the radius and $\tau/4m$ is the angular variable (recall that $\rho>0$ and that $\rho=0$ corresponds to $r=2m$). In polar coordinates $\rho=0$ is an apparent singularity iff the angle is identified with period $2\pi$, implying $\tau \sim \tau + 8\pi m$.
We have thus that the period of the imaginary time, $\beta_H = 8\pi m$, is the inverse of Hawking's temperature.
It is trivial to show that the metric is now smooth everywhere. Importantly, $r={\rm const.}$ are hypersurfaces with topology $S^1 \times S^2$, where $\tau$ is a coordinate on $S^1$, with radius increasing with $f(r)$, and $(\theta,\varphi)$ are coordinates on $S^2$.

If instead the angle is not identified with period $2\pi$, i.e., $\beta \neq \beta_H$, the Euclidean manifold will have a conical singularity.
A conical singularity can be seen as an infinitesimally small region of space with concentrated curvature. In the main paper we only need the integral of the scalar curvature, which is proportional to the deficit angle times the area of the codimension-two surface in which the singularity is located 
\begin{equation}
 \begin{split}
  \int_{{\rm sing}} \dd^4 x \sqrt{g} R
  =
  4\pi \Bigl( 1- \frac{\beta}{\beta_H}\Bigr) A \,,
 \end{split}
\end{equation}
where $A= 4\pi r_s^2 = 16 \pi m^2$ is the area of the horizon and the integral extends on an infinitesimal neighbor of $\rho=0$ \cite{Fursaev:1994te,Fursaev:1994ea,Fursaev:1995ef}


\end{document}